\documentclass[twocolumn,fp]{jpsj3}
\usepackage{graphicx}
\usepackage{bm}
\usepackage{cite}
\usepackage{color}
\usepackage{amssymb}
\usepackage{amsmath}
\usepackage{revsymb}

\title{Mott Transitions and Staggered Orders in the Three-component Fermionic System: Variational Cluster Approach}
\author{Takumi Hasunuma, Tatsuya Kaneko, Shohei Miyakoshi, and Yukinori Ohta\thanks{ohta@faculty.chiba-u.jp}}
\inst{Department of Physics, Chiba University, Chiba 263-8522, Japan}
\date{\today}

\abst{
The variational cluster approximation is used to study the ground-state properties and 
single-particle spectra of the three-component fermionic Hubbard model defined 
on the two-dimensional square lattice at half filling.  
First, we show that either a paired Mott state or color-selective Mott state is realized 
in the paramagnetic system, depending on the anisotropy in the interaction strengths, 
except around the SU(3) symmetric point, where a paramagnetic metallic state is 
maintained.  
Then, by introducing Weiss fields to observe spontaneous symmetry breakings, 
we show that either a color-density-wave state or color-selective antiferromagnetic 
state is realized depending on the interaction anisotropy and that the first-order phase 
transition between these two states occurs at the SU(3) point.  We moreover show 
that these staggered orders originate from the gain in potential energy (or Slater mechanism) 
near the SU(3) point but originate from the gain in kinetic energy (or Mott mechanism) 
when the interaction anisotropy is strong.  The staggered orders near the SU(3) point 
disappear when the next-nearest-neighbor hopping parameters are introduced, indicating 
that these orders are fragile, protected only by the Fermi surface nesting. 
}

\begin{document}
\maketitle

\section{Introduction}
The many-particle physics of multicomponent fermions in correlated electron 
systems, such as orbital orderings \cite{TN00}, 
orbital-selective Mott transitions \cite{KKRS04}, and 
orbital-spin liquids \cite{CLLPM12}, 
has recently been one of the major themes of condensed matter physics.  
The ground-state properties of the $N$-component 
model with $N>2$, where $N$ is the number of internal degrees of freedom of a fermion, 
are conjectured to depend strongly on $N$, and therefore the occurrence of novel 
quantum phases at $N>2$ has been anticipated \cite{AM88}.  
In particular, the SU($N$) symmetric model has been of great interest for simulating 
systems of correlated fermionic ultracold atoms in optical lattices 
\cite{Bl05,HGR09,CHU09,GHG10,NMCLT13}.  
Thus, the possible experimental realization of systems with $N>2$ will provide a good 
opportunity in the research field of quantum many-body problems, of 
which the candidates may be 
the SU(6) symmetry for $^\mathrm{173}$Yb \cite{FTK07,TYS12}, 
the SU(6)$\times$SU(2) symmetry for a mixture of $^\mathrm{171}$Yb and $^\mathrm{173}$Yb \cite{TTS10}, 
the SU(10) symmetry for $^\mathrm{87}$Sr \cite{DYM10,ZBB14}, and a three-component 
fermionic system, although not exactly SU(3), for $^\mathrm{6}$Li \cite{OLK08,HWH09}. 

The three-component fermionic Hubbard model, which is a natural extension of the 
two-component one for correlated electrons, has recently been studied to clarify its 
metal-insulator transition, translational-symmetry-broken staggered orders, and superfluidity.  
Namely, dynamical mean field theory (DMFT) calculations have shown that the metal-insulator 
transition in the paramagnetic state of the $N=3$ Hubbard model at half filling (or 3/2 
fermions per site) does not occur at the SU(3) symmetric point but it occurs when 
the interaction strengths between components are anisotropic \cite{GB09,IMS10,IS13}.  
Moreover, the DMFT calculations assuming translational symmetry breaking have 
shown that two types of staggered orders appear, depending on the anisotropy in the 
interaction strengths, and the first-order phase transition occurs between the two 
at the SU(3) symmetric point\cite{MIS10,IS13,YK16,YK161}.  
Superfluidity has also been reported to occur, even in the repulsive Hubbard model 
near half filling, when the interaction strengths are anisotropic, suggesting the presence 
of an exotic pairing mechanism \cite{IS12,IS13,OTK14,Su15}, where the fluctuations of the 
staggered orders may cause the pairing.  
We may therefore point out that the effects of the lattice geometry and Fermi surface 
nesting, which are important in the formation of the staggered orders, should be examined 
carefully in the low-dimensional systems.  

In this paper, motivated by the above developments in the field, we study the ground-state 
properties and excitation spectra of the three-component fermionic Hubbard model defined 
on the two-dimensional square lattice at half filling by means of the variational cluster 
approximation (VCA).  
The VCA can treat the short-range spatial correlations precisely in the thermodynamic 
limit of low-dimensional systems and therefore has a major advantage that the effects of 
the lattice geometry and Fermi surface topology can be examined, which DMFT studies 
cannot tackle.  
We thereby hope that some new insights into the physics of three-component fermionic 
systems will be obtained.  

First, we calculate the single-particle spectrum, density of states, and single-particle gap 
in the paramagnetic state to study the Mott transitions of the model, and we draw its 
ground-state phase diagram, which includes three distinct Mott phases.  
Then, we introduce the Weiss fields to study the translational symmetry breakings and 
show that two types of staggered orders appear in the ground state of the model.  
We moreover examine the origin of the long-range orderings in terms of the gains 
in kinetic and potential energies, addressing the Slater versus Mott mechanisms.  
We also examine the stability of the staggered orders by introducing the next-nearest-neighbor 
hopping parameters to destroy the Fermi surface nesting.  
We thereby discuss the characteristic properties of the three-component fermionic 
system defined on the two-dimensional square lattice.  

This paper is organized as follows.  
In Sect.~2, we introduce the three-component Hubbard model and briefly summarize the method of VCA.  
In Sect.~3, we present calculated results for the paramagnetic and staggered ordered states and give 
some discussion.  A summary of the paper is given in Sect.~4. 
 
\section{Model and Method}

\subsection{Three-component Hubbard model}

The three-component Hubbard model may be defined by the Hamiltonian 
\begin{align}
\mathcal{H}=& -t \sum_{\langle i,j \rangle}\sum_{\alpha}  c_{i\alpha}^{\dagger}c_{j\alpha}
-\sum_i\sum_{\alpha}\mu_{\alpha}n_{i\alpha} \notag \\
&+\frac{1}{2}\sum_i\sum_{\alpha\ne\beta}U_{\alpha\beta}n_{i\alpha}n_{i\beta}, 
\label{ham}
\end{align} 
where $c^{\dag}_{i\alpha}$ ($c_{i\alpha}$) denotes the creation (annihilation) operator 
of a fermion with color 
$\alpha$ $(=a, b, c)$ at site $i$ and $n_{i\alpha} = c^\dag_{i\alpha}c_{i\alpha}$.  
$t$ is the hopping integral between the neighboring sites, which is taken as the unit of energy, 
and $U_{\alpha\beta}$ $(=U_{\beta\alpha})$ is the on-site interaction between two fermions 
with colors $\alpha$ and $\beta$.  
Throughout the paper, we assume the color-dependent interactions and set 
$U_{ab}=U$ $(>0)$ and $U_{bc}=U_{ca}=U'$ $(>0)$ for simplicity.  
We also assume the filling of $n=\sum_{\alpha}n_{\alpha}=3/2$ (denoted as half filling), 
where $n_{\alpha}\equiv \sum_i \langle n_{i\alpha} \rangle/L = 1/2$ is the average number 
of color-$\alpha$ fermions in a system of size $L$.  
We set the chemical potential as $\mu_{\alpha} = \sum_{\beta\ne\alpha}U_{\alpha\beta}/2$ 
to maintain the average particle density at $n=3/2$.  

This model with $U=U'$ corresponds to the SU(3) Hubbard model, using which studies have 
been carried out on superfluidity in the presence of attractive interactions as well as on the 
metal-insulator transition in the presence of repulsive interactions 
\cite{HH04,HH04-2,RZH07,RHZ08,IS09,SK14}.  
In particular, it has been confirmed \cite{GB09} that this model at half filling does 
not show the Mott transition in the paramagnetic state, maintaining the metallic state 
irrespective of the interaction strength, unlike in the two-component Hubbard model.  
At $U'=0$, on the other hand, the two interacting components undergo the Mott transition, 
leaving the noninteracting component metallic.   
Thus, the anisotropy in the interaction strengths ($U\ne U'$) causes the Mott transition: 
at $U \gg U'$, the color-selective Mott (CSM) state is realized, where the two components 
are localized and one component is itinerant, and at $U \ll U'$, the paired Mott (PM) 
state is realized, where the two components are paired in the same sites and one 
component is localized in the other sites \cite{IMS10,IS13,OTK14}.  

In the presence of long-range staggered orders, it is known that fermions are arranged 
alternately in the lattice and either the color-selective antiferromagnetic (CSAF) state 
corresponding to the CSM state at $U \gg U'$ or the color-density-wave (CDW) state 
corresponding to the PM state at $U \ll U'$ is realized \cite{MIS10,IS13,YK16}.  
It has been pointed out \cite{HH04} that in the SU(3) symmetric $U=U'$ Hubbard model, 
the CDW state is realized at half filling if the perfect Fermi surface nesting of the nesting 
vector $\bm{Q}=(\pi,\pi)$ exists in the two-dimensional square lattice.  
It has also been pointed out \cite{MIS10,IS13} that in the SU(3) symmetric model at 
half filling, the CSAF and CDW states are energetically degenerate at zero temperature 
and that the CSAF state is realized at $U > U'$ and the CDW state is realized at $U < U'$.  
DMFT calculations have suggested the presence of an $s$-wave superfluid state 
at $U'>U>0$, which is, however, higher in energy than the CDW state and is not realized 
as the ground state of the system \cite{OTK14,YK16}.  

\subsection{Variational cluster approximation}

To accomplish the calculations in the thermodynamic limit, we use the VCA \cite{PAD03,Po12} 
based on self-energy functional theory (SFT), which is the variational principle for 
the grand potential as a functional of the self-energy \cite{Po03-1,Po03-2,Po12}.  Unlike in 
DMFT, we can thereby precisely take into account the effects of short-range spatial fermionic 
correlations in low-dimensional systems.  
In fact, successful explanations were given for the antiferromagnetism and 
superconductivity \cite{SLM05}, as well as for the pseudogap behaviors \cite{ST04}, in the 
two-dimensional Hubbard model for high-$T_c$ cuprate materials.  
The trial self-energy for the variational method is generated from the 
exact self-energy of the disconnected finite-size clusters, which act as a reference system.  
To investigate the spontaneous symmetry breaking in the VCA~\cite{DAH04}, we introduce 
the Weiss fields in the system as variational parameters.  
The Weiss fields of the CDW and CSAF states are defined as 
\begin{align}
\mathcal{H}'_{\mathrm{CDW}}&=M'_{\mathrm{CDW}} \sum_{i} e^{i\bm{Q}\cdot\bm{r}_i}\left( n_{ia}+n_{ib} - n_{ic} \right) 
\label{WFCDW} \\
\mathcal{H}'_{\mathrm{CSAF}}&=M'_{\mathrm{CSAF}} \sum_{i} e^{i\bm{Q}\cdot\bm{r}_i}\left( n_{ia}-n_{ib}\right),
\label{WFCSAF}
\end{align} 
respectively, where $M'_{\mathrm{CDW}}$ and $M'_{\mathrm{CSAF}}$ are the strengths 
of the Weiss fields of the CDW and CSAF states, respectively, which are taken as the variational 
parameters.  Then, the Hamiltonian of the reference system is given by 
$\mathcal{H}'=\mathcal{H}+\mathcal{H}'_{\mathrm{CDW}}+\mathcal{H}'_{\mathrm{CSAF}}$. 
Within SFT, the grand potential at zero temperature is given by 
\begin{align}
\Omega=\Omega'-\frac{1}{N_s}\oint_{C}\frac{{\rm d}z}{2\pi i}\sum_{\bm{K},\alpha}\ln\det
\left[\bm{I}-\bm{V}_{\alpha}(\bm{K})\bm{G}'_{\alpha}(z)\right], 
\end{align}
where $\Omega'$ is the grand potential of the reference system, $N_s$ is the number of clusters 
in the system, $\bm{I}$ is the unit matrix, $\bm{V}_{\alpha}$ is the hopping parameter 
between the adjacent clusters, and $\bm{G}'_{\alpha}$ is the exact Green's function of 
the reference system calculated by the Lanczos exact-diagonalization method.  
The $\bm{K}$-summation is performed in the reduced Brillouin zone of the superlattice 
and the contour $C$ of the frequency integral encloses the negative real axis.  
The variational parameters are optimized on the basis of the variational principle, i.e., 
$\partial\Omega/\partial M'_{\mathrm{CDW}}=0$ for the CDW state and 
$\partial\Omega/\partial M'_{\mathrm{CSAF}}=0$ for the CSAF state.  
The solutions with $M'_{\mathrm{CDW} }\ne 0$ and $M'_{\mathrm{CSAF}}\ne 0$ correspond 
to the CDW and CSAF states, respectively.  
In our VCA calculations, we assume the two-dimensional square lattice and the modulation vector 
is given by $\bm{Q}=(\pi,\pi)$.  We use an $L_c=2\times 3$ site cluster as the reference system.  

To calculate the single-particle spectrum and density of states (DOS), we use cluster perturbation 
theory (CPT) \cite{SPP00,SPP02,Se12}, which proceeds by tiling the lattice into identical, finite-size 
clusters, solving many-body problems in these clusters exactly, and treating the intercluster 
hopping terms at the first order in strong-coupling perturbation theory.  This theory is exact in 
both the strong and weak correlation limits, and provides a good approximation to the spectral 
function at any wave vector.  
In CPT, the Green's function of color-$\alpha$ fermions is given by 
\begin{align}
\mathcal{G}^{\mathrm{cpt}}_{\alpha}(\bm{k},\omega)
=\frac{1}{L_c}\sum^{L_c}_{i,j=1}\mathcal{G}_{ij,\alpha}(\bm{k},\omega)e^{-i\bm{k}\cdot(\bm{r}_i-\bm{r}_j)},
\end{align}
where  $\bm{\mathcal{G}}_{\alpha}(\bm{K},\omega)=\big[\bm{G}'^{-1}_{\alpha}(\omega)-\bm{V}_{\alpha}(\bm{K})\big]^{-1}$.
Using the CPT Green's function $\mathcal{G}^{\mathrm{cpt}}_{\alpha}$, the single-particle 
spectral function of the color-$\alpha$ fermions is defined as 
\begin{align}
A_{\alpha}(\bm{k},\omega)=-\frac{1}{\pi}\mathrm{Im}\; \mathcal{G}^{\mathrm{cpt}}_{\alpha}(\bm{k},\omega+i\eta), 
\label{Akw}
\end{align}
where $\eta$ gives the artificial Lorentzian broadening of the spectrum.  
We also calculate the DOS of the color-$\alpha$ fermions defined as
\begin{align}
\rho_{\alpha}(\omega)=\frac{1}{L}\sum_{\bm{k}}A_{\alpha}(\bm{k},\omega), 
\end{align}
where $L=N_sL_c$ is the total number of lattice sites in the system.  

\begin{figure*}[htb]
\begin{center}
\includegraphics[width=2.04\columnwidth]{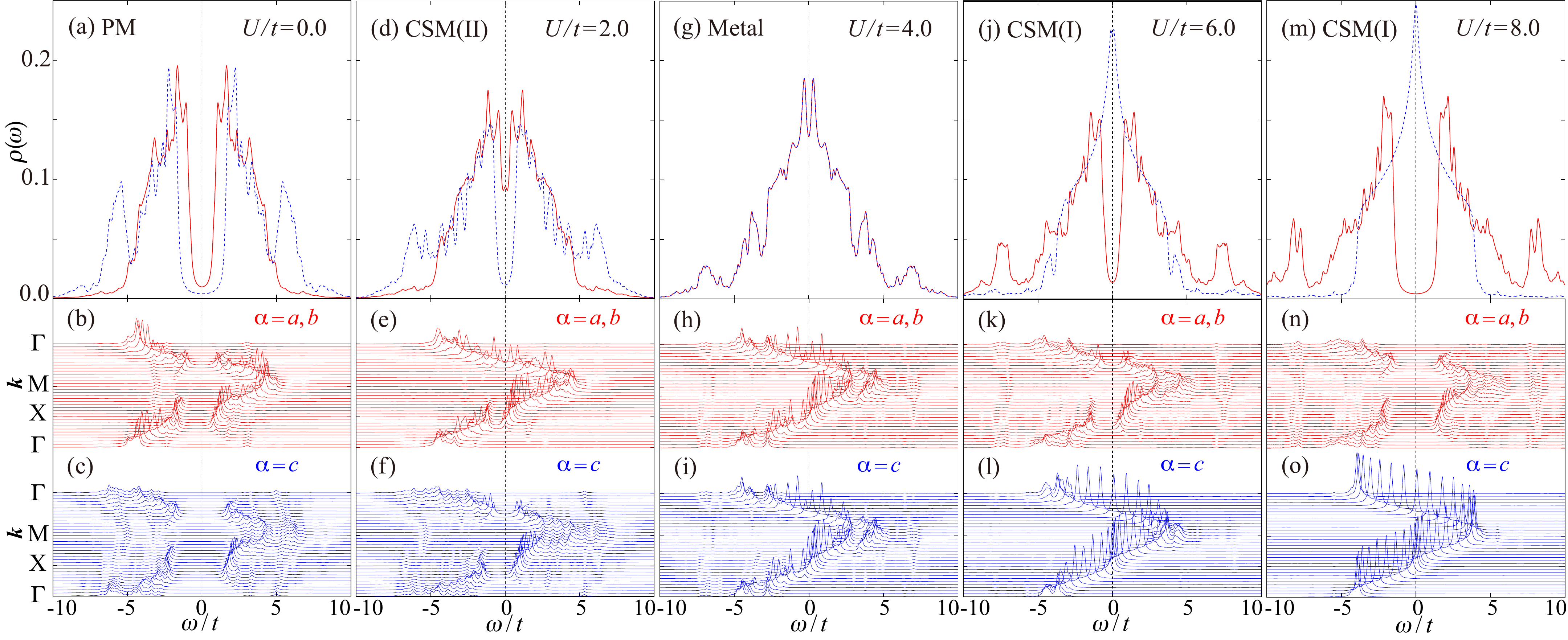}
\caption{(Color online) 
Calculated DOSs and single-particle spectra in the normal state of our model [Eq.~(\ref{ham})].  
The $U/t$ dependence at $U'/t=4$ is shown, where the solid (dotted) lines indicate the results for 
the color $a$ and $b$ (color $c$) components.  The dispersions of the spectra are shown along 
the line connecting the $\Gamma(0,0)$, X$(\pi,0)$, and M$(\pi,\pi)$ points of the Brillouin zone.  
The artificial Lorentzian broadening of the spectra $\eta/t=0.1$ is used.  
}\label{fig1}
\end{center}
\end{figure*}

\section{Results of Calculations}

\subsection{Mott transitions in the paramagnetic state}

\begin{figure}[htb]
\begin{center}
\includegraphics[width=\columnwidth]{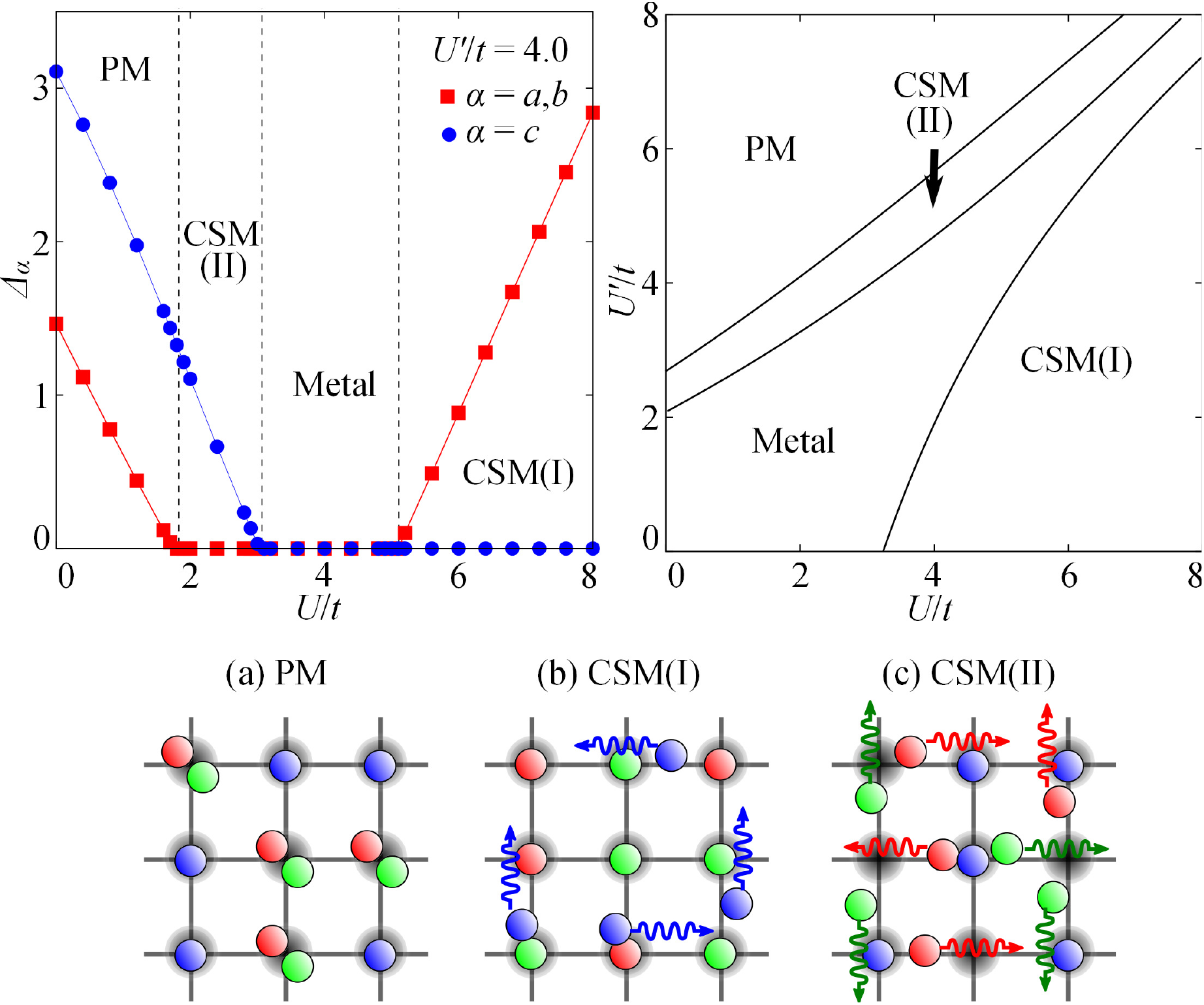}
\caption{(Color online) 
Left panel: Calculated single-particle gap $\varDelta_\alpha$ for color $\alpha$ fermions as a function of $U/t$ at $U'/t=4$.  
Right panel: Calculated phase diagram of the paramagnetic state in the parameter space $(U/t,U'/t)$, which includes 
the paired Mott (PM) and two types of color-selective Mott (CSM) phases.  
Illustrated below are their schematic representations.  
}\label{fig2}
\end{center}
\end{figure}

First, let us discuss the Mott metal-insulator transitions in the paramagnetic state 
(or in the absence of the Weiss fields).  Using the CPT, we calculate the single-particle 
spectra and DOS, the results of which are shown in Fig.~\ref{fig1} as a function of 
$U/t$ at $U'/t=4$.  
At $U/t=0$, where the PM state is realized [see Fig.~\ref{fig2}(a)], we find that the gap 
opens in all components of the spectra at the Fermi level; 
in particular, the gap in the $c$ component is about twice as large as the gap in the 
$a$ and $b$ components.  
This may be understood as follows: if a color $a$ (or $b$) fermion is added to the 
system, it is placed on the site occupied by the color $c$ fermion due to the Pauli principle, 
which increases the energy of the system by $U'$, but if a color $c$ fermion is added 
to the system, it is placed on the site occupied doubly by the color $a$ and $b$ fermions, 
which increases the energy of the system by $2U'$.  Thus, a gap of size $U'$ ($2U'$) 
opens in the $a$ and $b$ components ($c$ component) of the spectra.  

With increasing $U/t$, we find that the gap in the $a$ and $b$ components decreases 
and closes at a certain $U/t$ value, but the gap in the $c$ component remains open 
up to a larger $U/t$ value.  
Then, at $U/t=U'/t=4$, where our model is SU(3) symmetric, the spectra become 
equivalent in all the components and the system becomes metallic as shown in 
Fig.~\ref{fig1}(g).  

Increasing the $U/t$ value further, we find that a gap opens again in the $a$ and $b$ 
components, but the $c$ component remains metallic and the DOS curve 
resembles that of the noninteracting band in the two-dimensional square lattice.  
This situation occurs because the color $a$ and $b$ fermions are localized owing to 
the large $U/t$ values, whereas the color $c$ fermions hop freely because the interaction 
strengths are the same, $U_{bc}=U_{ca}=U'$.   
In this CSM state, a gap of size $U$ opens in the $a$ and $b$ components of 
the spectra because, if a color $a$ ($b$) fermion is added to the system, it is placed 
on the site occupied by the color $b$ ($a$) fermion due to the Pauli principle, 
which increases the energy of the system by $U$.  

Note that the size of the single-particle gap cannot be estimated accurately in 
Fig.~\ref{fig1} because the spectra are broadened artificially by $\eta$ [see Eq.~(\ref{Akw})].  
However, the chemical-potential dependence of the average particle number per site 
$n_{\alpha}$ enables us to evaluate the gap size accurately, which may be 
calculated as 
\begin{align}
n_{\alpha}=\frac{1}{L}\sum_{i=1}^{L}\langle n_{i\alpha} \rangle
=\frac{1}{N_sL_c}\oint_{C}\frac{{\rm d}z}{2\pi i}\sum_{\bm{K}}\sum_{i=1}^{L_c}\mathcal{G}_{ii,\alpha}(\bm{K},z) 
\end{align}
via the diagonal term of the Green's function $\bm{\mathcal{G}}_{\alpha}$.  
The particle number $n_{\alpha}$ calculated as a function of the chemical potential 
$\mu_{\alpha}$ is fixed to $n_{\alpha}=0.5$ at half filling if the system is metallic, but 
it shows a plateau in the range $\mu_{\alpha,-} < \mu_{\alpha} < \mu_{\alpha,+}$, 
where $\mu_{\alpha,+}$ corresponds to the lower edge of the upper band and 
$\mu_{\alpha,-}$ corresponds to the upper edge of the lower band.  The width 
of the plateau is then given by $\varDelta_{\alpha}=|{\mu_{\alpha,+}-\mu_{\alpha,-}}|$, 
which corresponds to the single-particle gap.  

The thus calculated single-particle gaps $\varDelta_{\alpha}$ are shown in 
Fig.~\ref{fig2} as a function of $U/t$ at $U'/t=4$.  We find that at $U/t=0$, a gap opens 
in all components of the spectra ($\varDelta_{\alpha}>0$), as shown in Figs.~\ref{fig1}(a)-\ref{fig1}(c), 
which is in accordance with the PM state previously discussed \cite{IMS10,IS13}.  
The sizes of the gaps are $U'$ in the $a$ and $b$ components and $2U'$ in the $c$ 
component, as we have discussed above, so that we obtain the relation 
$\varDelta_{c}\simeq 2\varDelta_{a}=2\varDelta_{b}$ at $U/t=0$.  
With increasing $U/t$, the gaps $\varDelta_{\alpha}$ decrease linearly, and in the region 
corresponding to Figs.~\ref{fig1}(d)-\ref{fig1}(f), we have $\varDelta_{a}=\varDelta_{b}=0$ and 
$\varDelta_{c}>0$, the region of which we call CSM(II), where the color $a$ and $b$ 
fermions have metallic behavior and the color $c$ fermions have insulating behavior.  
Upon increasing $U/t$ further, the gap in the $c$ component also closes, and around 
$U=U'$, the gaps in all the components close ($\varDelta_{\alpha}=0$), as shown 
in Figs.~\ref{fig1}(g)-\ref{fig1}(i).  
In the large-$U/t$ $(\gg U'/t)$ region, we have $\varDelta_{a}=\varDelta_{b}>0$ and $\varDelta_{c}=0$, 
the state of which we call CSM(I), where the gaps $\varDelta_{\alpha}$ ($\alpha=a, b$) increase 
linearly with $U$, indicating that the CSM(I) phase is caused by $U$.  
The critical phase boundaries determined at $U'/t=4$ are 
$U/t\le 1.8$ for PM, 
$1.8\le U/t\le 3.1$ for CSM(II), and
$5.1\le U/t$ for CSM(I).  

Figure \ref{fig2} shows the phase diagram determined by the thus calculated gaps 
$\varDelta_{\alpha}$, where we find three Mott phases, PM, CSM(I), and CSM(II), as well as 
the paramagnetic metallic phase.  The CSM(I) phase appears at $U\gg U'$ and the PM phase 
appears at $U\ll U'$, whereas around the SU(3) symmetric point $U=U'$, we find the paramagnetic 
metallic phase.  The newly found CSM(II) phase appears between the PM and paramagnetic 
metallic phases, where only the color $c$ fermions are gapful.  This phase appears because 
the single-particle excitation requires twice the energy for the $c$ component as for the 
$a$ and $b$ components, as shown above.  
This phase is absent in the DMFT calculations \cite{IMS10,IS13}, suggesting its absence in 
the case of infinite dimensions; thus, we consider that the intersite spatial correlations 
between fermions in two-dimension, which the VCA takes into account properly, may induce 
this CSM(II) phase.  

\begin{figure}[htb]
\begin{center}
\includegraphics[width=\columnwidth]{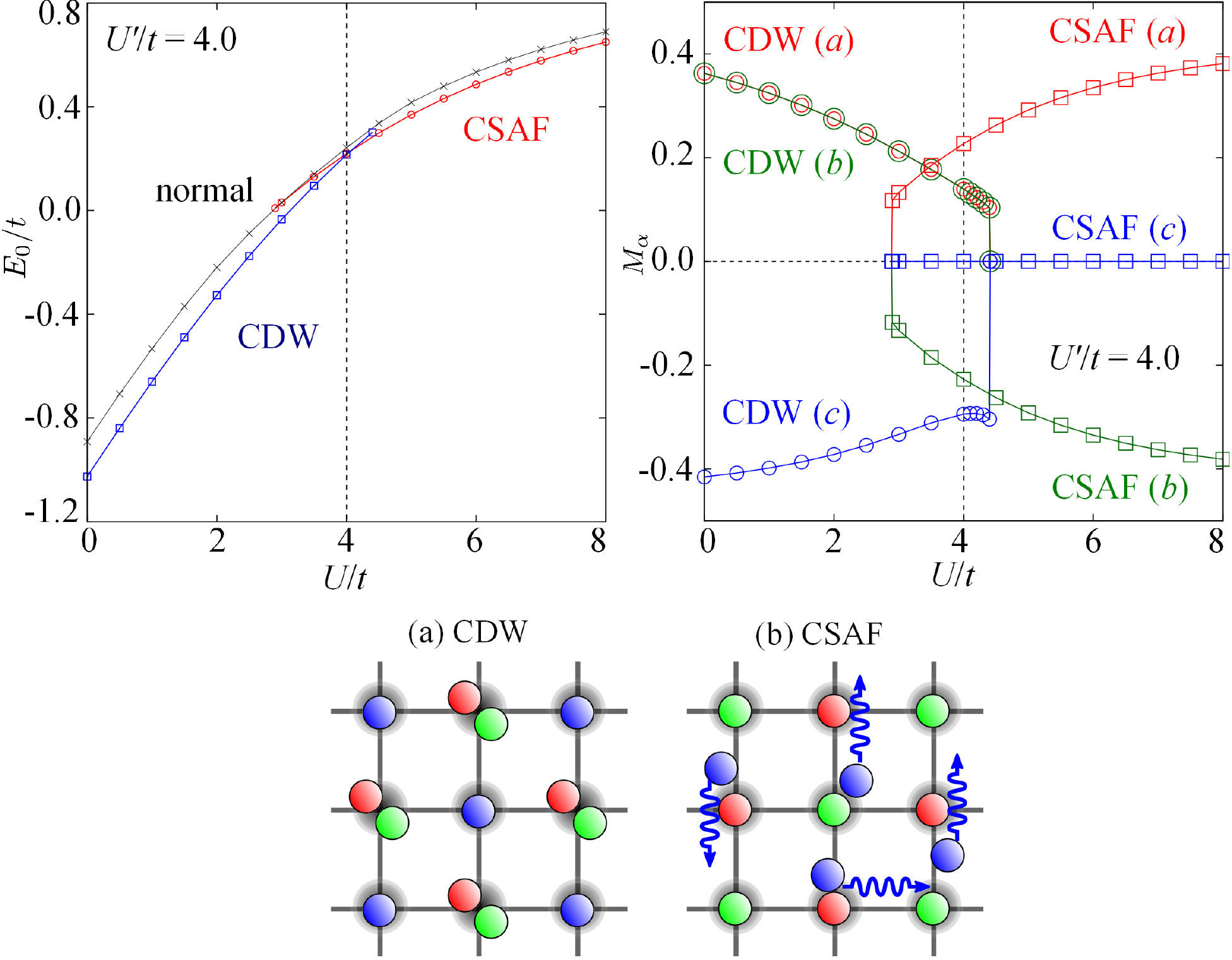}
\caption{(Color online) 
Calculated ground-state energies $E_0/t$ (left panel) and staggered magnetizations 
$M_{\alpha}$ (right panel) of the color-density-wave (CDW) and color-selective 
antiferromagnetic (CSAF) states as a function of $U/t$ at $U'/t=4$.  
Illustrated below are schematic representations of the CDW and CSAF states.  
}\label{fig3}
\end{center}
\end{figure}

\begin{figure}[htb]
\begin{center}
\includegraphics[width=\columnwidth]{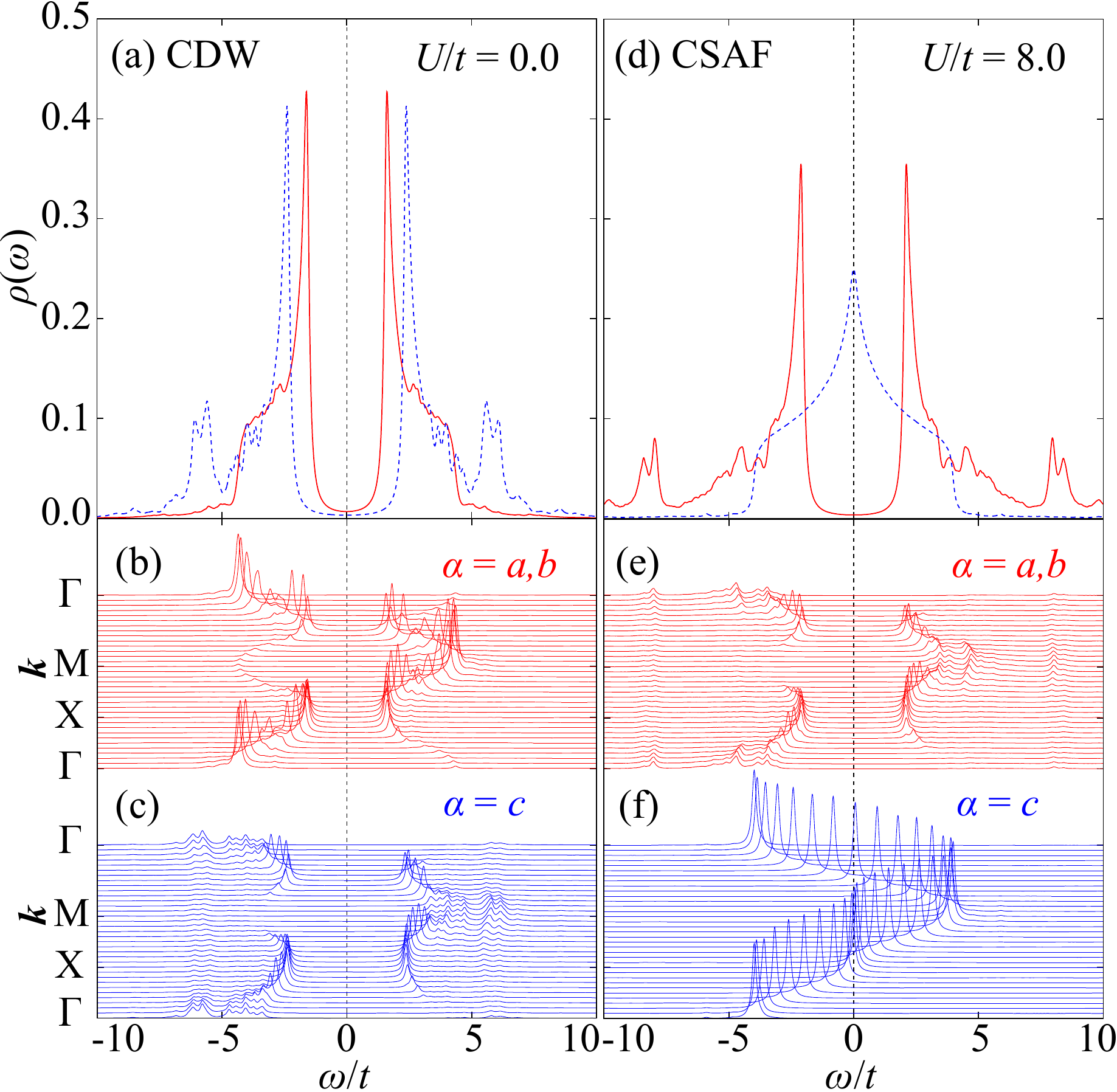}
\caption{(Color online) 
Calculated DOSs and single-particle spectra in the CDW (at $U/t=0$ and $U'/t=4$) and 
CSAF (at $U/t=8$ and $U'/t=4$) states.  The solid (dotted) curves indicate the spectra 
for color $a$ and $b$ fermions (color $c$ fermions).  
The dispersions of the spectra are shown along the line connecting the $\Gamma(0,0)$, 
X$(\pi,0)$, and M$(\pi,\pi)$ points of the Brillouin zone.  
The artificial Lorentzian broadening of the spectra $\eta/t=0.1$ is used.  
}\label{fig4}
\end{center}
\end{figure}

\subsection{Staggered ordered states}

Next, let us discuss the symmetry-broken staggered ordered states, which are 
obtained by adding the Weiss fields defined in Eqs.~(\ref{WFCDW}) and (\ref{WFCSAF}) 
to the Hamiltonian in Eq.~(\ref{ham}).  We thereby calculate the ground-state energy per 
site, $E_0=\Omega+\sum_{\alpha}\mu_{\alpha}$, and the staggered magnetization of 
color $\alpha$ fermions defined as 
\begin{align}
\label{7}
M_{\alpha}&= \frac{1}{L} \sum_{i}  \langle n_{i\alpha} \rangle e^{i\bm{Q}\cdot\bm{r}_i} \notag \\
&=\frac{1}{N_sL_c}\oint_{C}\frac{{\rm d}z}{2\pi i}\sum_{\bm{K}}\sum^{L_c}_{i=1}
\mathcal{G}_{ii,\alpha}(\bm{K},z) e^{i\bm{Q}\cdot\bm{r}_i} 
\end{align}
for the optimized Weiss fields.  

The calculated results for $E_0$ and $M_{\alpha}$ are shown in Fig.~\ref{fig3} as a 
function of $U$ at $U'/t=4$.  We find that the ground-state energies $E_0$ of the 
CDW and CSAF states cross each other at $U=U'$, indicating that the phase transition 
between the two is of the first order.  The CDW (CSAF) is thus realized as 
the ground state when $U'$ ($U$) is larger than $U$ ($U'$), in accordance with previous 
DMFT studies \cite{MIS10,IS13,YK16}.  
Around the SU(3) symmetric point (or around $U\simeq U'$), the grand potential 
$\Omega$ has stationary points both at $M'_{\mathrm{CDW}}\ne0$ and at 
$M'_{\mathrm{CSAF}}\ne0$, indicating that either the CDW or CSAF phase can appear.  
The calculated staggered magnetization indicates that $M'_{\mathrm{CDW}}\ne 0$ at 
$U<U'$, where the CDW state is stable.  In particular, we find that $M_a=M_b>0$ and 
$M_c<0$, which indicate that the color $a$ and $b$ fermions are located on the 
same sites and the color $c$ fermions are located alternately on other sites, 
resulting in the staggered order of the modulation vector $\bm{Q}=(\pi,\pi)$.  
At $U>U'$, we find that $M'_{\mathrm{CSAF}}\ne 0$, where the CSAF state is stable.  
In particular, we find that $M_a>0$, $M_b<0$, and $M_c=0$, which indicate that the 
color $a$ and $b$ fermions show staggered antiferromagnetic orderings but 
the color $c$ fermions do not.  

We also calculate the single-particle spectra and DOSs for the staggered ordered 
CDW and CSAF phases, where we use the optimized Weiss fields $M'_{\mathrm{CDW}}$ 
and $M'_{\mathrm{CSAF}}$.  The results are shown in Fig.~\ref{fig4}, where we 
find that the spectral peak positions do not change markedly in comparison with those 
of the corresponding PM and CSM(I) phases, whereas sharp coherence peaks appear 
at the edges of the gap in both the CDW and CSAF phases.  These gaps become 
wider than those of the corresponding PM and CSM(I) phases, reflecting the stabilization 
of the ordered phases.  

\begin{figure}[thb]
\begin{center}
\includegraphics[width=\columnwidth]{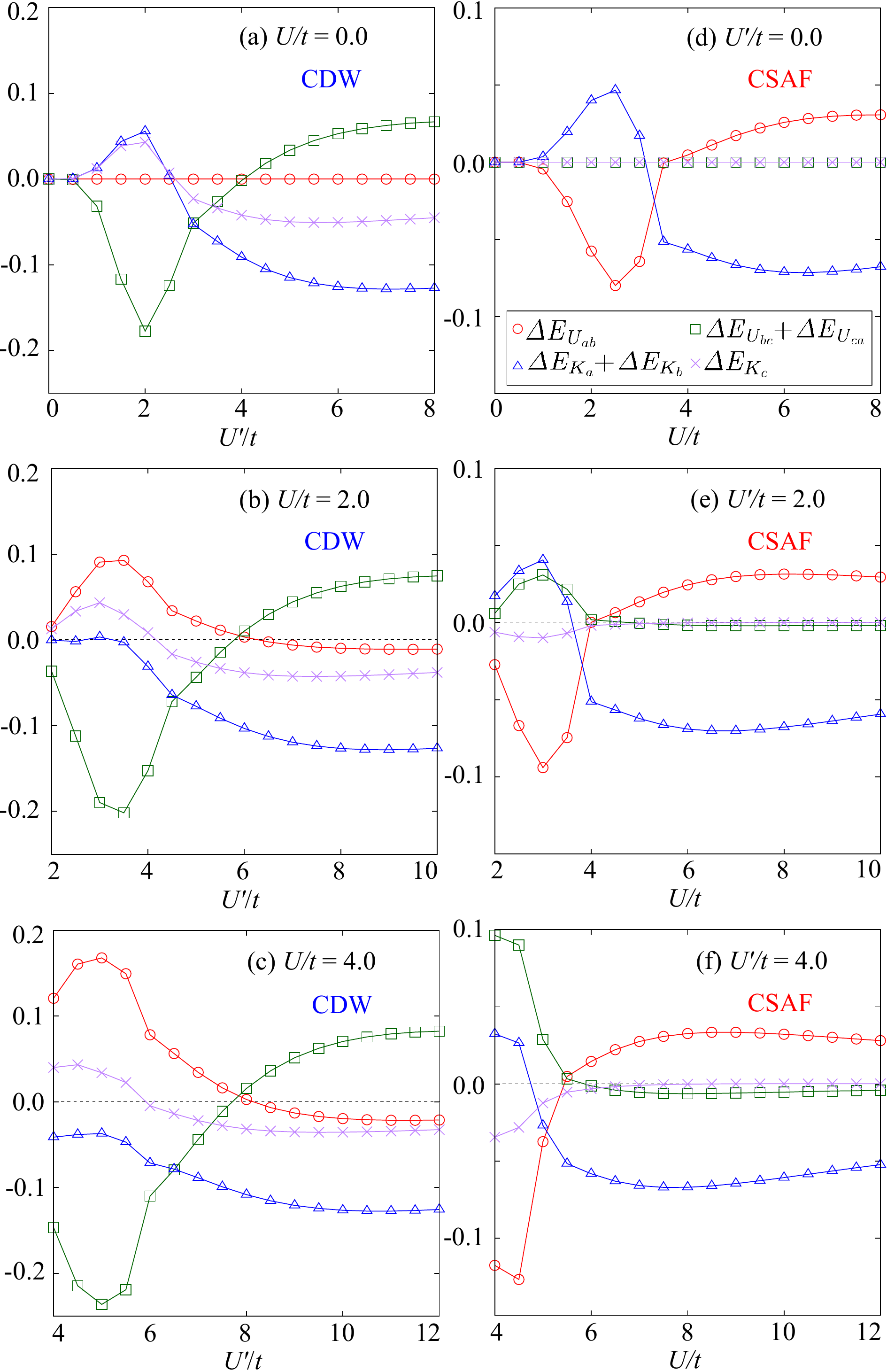}
\caption{(Color online) 
Calculated kinetic and potential energy differences in the CDW (left panels) and CSAF (right panels) 
phases compared with the paramagnetic normal phase.  
Plotted are $\varDelta E_{U_{ab}}$ ($\circ$), 
$\varDelta E_{U_{bc}}+\varDelta E_{U_{ca}}$ ($\square$),  
$\varDelta E_{K_{a}}+\varDelta E_{K_{b}}$ ($\triangle$), and 
$\varDelta E_{K_{c}}$ ($\times$) in units of $t$.  
$U/t$ and $U'/t$ dependences are shown.  
}\label{fig5}
\end{center}
\end{figure}

\begin{figure}[thb]
\begin{center}
\includegraphics[width=\columnwidth]{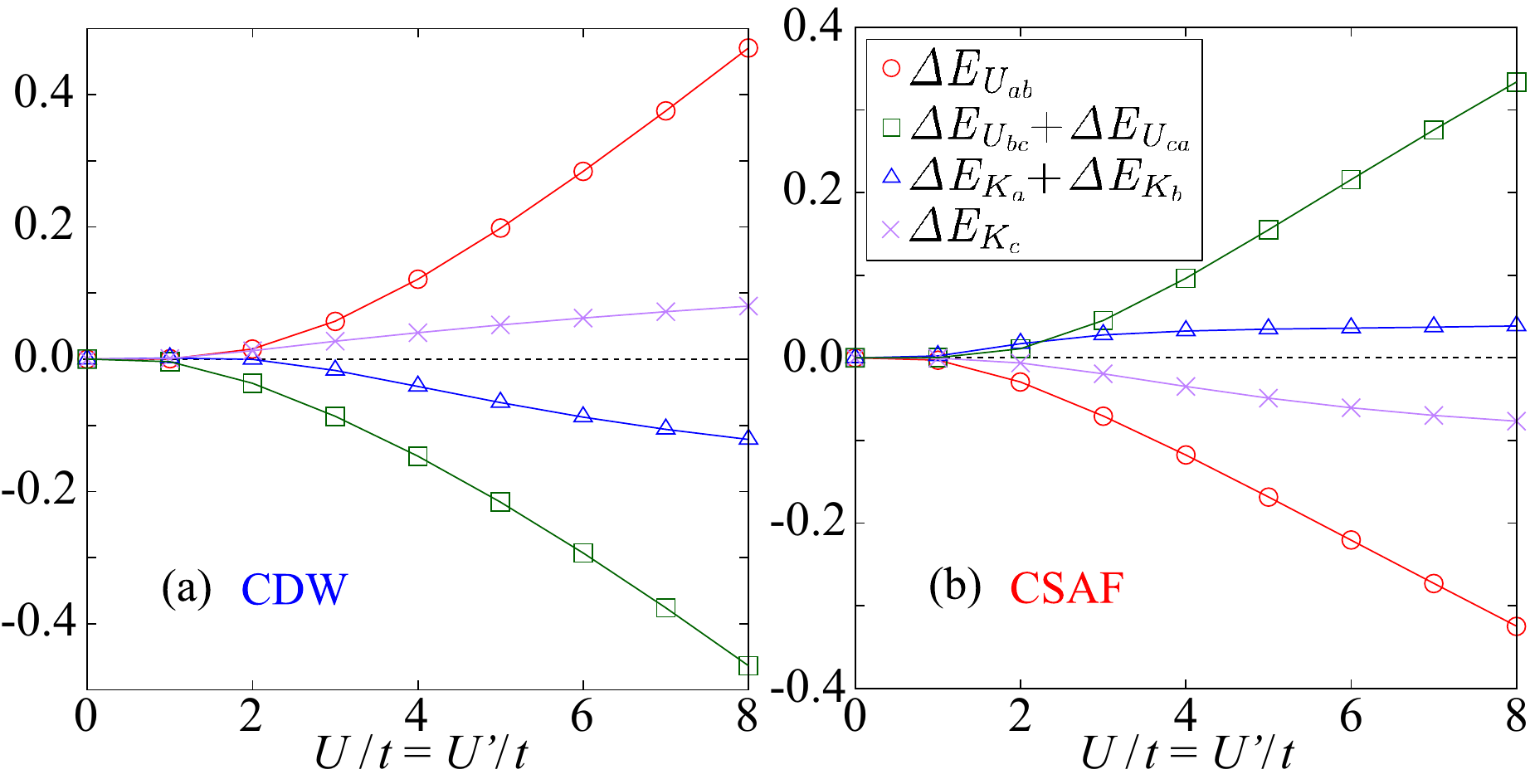}
\caption{(Color online) 
Same as in Fig.~\ref{fig5} but for the dependence on $U/t=U'/t$.  
}\label{fig6}
\end{center}
\end{figure}

\subsection{Slater versus Mott mechanisms}

In the previous subsection, we showed that the CDW and CSAF phases are stabilized for 
$U<U'$ and $U>U'$, respectively, and that the first-order phase transition occurs between 
the two phases at $U=U'$.  If we assume the paramagnetic phase, the Mott insulating 
phases such as PM and CSM are stabilized when the interaction strengths are strongly 
anisotropic (see Fig.~\ref{fig2}), but the paramagnetic metallic phase is maintained around 
the SU(3) symmetric point ($U=U'$) even if the interactions are very strong.  
Therefore, we may anticipate that the mechanisms of the stabilization of the staggered 
orders are different in two regions: the region around $U=U'$ where the system is 
metallic and the region where the interaction strengths are strongly anisotropic and the 
system is Mott insulating.  

It is known that there are two mechanisms for the stabilization of staggered orders 
\cite{YOT13,KO14,HSY14}.  
One is the Slater mechanism, which occurs in itinerant systems, where the Fermi surface 
instability causes band folding and gap opening due to the interactions between 
fermions, leading to staggered orderings in the system, resulting in gains in the potential 
energy but losses in the kinetic energy.  
The other is the Mott mechanism, which occurs in insulating systems, where the virtual 
hopping of fermions aligns their colors antiferromagnetically due to the Pauli principle, 
leading to staggered orderings in the system, resulting in gains in the kinetic energy but 
losses in the potential energy.  
Below, we calculate the kinetic and potential energies in the paramagnetic and staggered 
ordered phases of the system, and consider the mechanisms of the staggered orderings 
from the energetic point of view.  

Let us define the potential $E_{U_{\alpha\beta}}$ and kinetic $E_{K_{\alpha}}$ energies per site as
\begin{align}
E_{U_{\alpha\beta}}&=\frac{1}{L}\Biggl( U_{\alpha\beta}  \sum_{i}{\langle n_{i\alpha}n_{i\beta}\rangle}\Biggr)=U_{\alpha\beta}\frac{\partial E_{0}}{\partial{U_{\alpha\beta}}} \\
E_{K_{\alpha}}&=\frac{1}{L} \Biggl( -t_{\alpha} {\sum_{\langle i,j \rangle }{\langle c_{i\alpha}^{\dagger}c_{j\alpha} \rangle}}\Biggr)=-t_{\alpha}\frac{\partial E_{0}}{\partial{t_{\alpha}}}, 
\end{align}
and their energy gains caused by the staggered orderings as 
\begin{align}
\it{\Delta} E_{U_{\alpha\beta}}^{\mathrm{DW}}&=E_{U_{\alpha\beta}}^{\mathrm{DW}}-E_{U_{\alpha\beta}}^{\mathrm{N}} \\
\it{\Delta} E_{K_{\alpha}}^{\mathrm{DW}}&=E_{K_{\alpha}}^{\mathrm{DW}}-E_{K_{\alpha}}^{\mathrm{N}}
\end{align}
for the CDW phase at $U\leq U'$, and as 
\begin{align}
\it{\Delta} E_{U_{\alpha\beta}}^{\mathrm{AF}}&=E_{U_{\alpha\beta}}^{\mathrm{AF}}-E_{U_{\alpha\beta}}^{\mathrm{N}} \\
\it{\Delta} E_{K_{\alpha}}^{\mathrm{AF}}&=E_{K_{\alpha}}^{\mathrm{AF}}-E_{K_{\alpha}}^{\mathrm{N}}
\end{align}
for the CSAF phase at $U\geq U'$, 
where the superscripts DW, AF, and N stand for the CDW, CSAF, and paramagnetic normal phases, respectively.  
Thus, comparing the signs and magnitudes of $\it{\Delta} E_{U_{\alpha\beta}}$ and $\it{\Delta} E_{K_{\alpha}}$, 
we can evaluate the energy gains in the formation of the staggered long-range orders.   

Figure \ref{fig5} displays the calculated results for the quantities defined above as a function of 
$U/t$ $(U'/t)$ at a fixed value of $U'/t$ $(U/t)$.  
The same results along the line $U/t=U'/t$ are also shown in Fig.~\ref{fig6}.  
First, in Figs.~\ref{fig5}(a)-\ref{fig5}(c), where a comparison is made between the CDW and normal phases, 
we find that 
\begin{align}
&\varDelta E_{U_{bc}}^{\mathrm{DW}}+\varDelta E_{U_{ca}}^{\mathrm{DW}}<0 \\
&\varDelta E_{K_{a}}^{\mathrm{DW}}+\varDelta E_{K_{b}}^{\mathrm{DW}}>0 \\
&\varDelta E_{K_{c}}^{\mathrm{DW}}>0
\end{align}
for $0.0 \leq U'/t < 2.0$ at $U/t=0$, indicating that the gain in potential energy leads 
to the staggered CDW order.  
For $4.0 \leq U'/t \leq 8.0$, however, the signs are inverted and we find that 
\begin{align}
&\varDelta E_{U_{bc}}^{\mathrm{DW}}+\varDelta E_{U_{ca}}^{\mathrm{DW}}>0 \\
&\varDelta E_{K_{a}}^{\mathrm{DW}}+\varDelta E_{K_{b}}^{\mathrm{DW}}<0 \\
&\varDelta E_{K_{c}}^{\mathrm{DW}}<0,
\end{align}
which indicates that the gain in kinetic energy leads to the staggered order.  
Thus, the stabilization mechanism of the CDW phase shows a crossover from the 
Slater mechanism to the Mott mechanism.  
With increasing $U/t$, the interaction between the $a$ and $b$ components increases and 
$\varDelta E_{U_{ab}}^{\mathrm{DW}}$ varies considerably as shown in Figs.~\ref{fig5}(b) and \ref{fig5}(c).  
We then find that $\varDelta E_{U_{ab}}^{\mathrm{DW}}>0$ at $U\simeq U'$ and 
$\varDelta E_{U_{ab}}^{\mathrm{DW}}<0$ at $U' \gg U$.  
We also find that around $U=U'$, although the loss in the potential energy is large, 
$\varDelta E_{U_{ab}}^{\mathrm{DW}}>0$, the effect of 
$\varDelta E_{U_{bc}}^{\mathrm{DW}}+\varDelta E_{U_{ca}}^{\mathrm{DW}}<0$ is still dominant, 
leading to the Slater mechanism of the CDW ordering.  
This mechanism even occurs at large $U=U'$ values [see Fig.~\ref{fig6}(a)], 
suggesting that the effects of Fermi surface nesting are important here.  

Next, in Figs.~\ref{fig5}(d)-\ref{fig5}(f) where a comparison is made between the CSAF and normal phases, 
we find that 
\begin{align}
&\varDelta E_{U_{ab}}^{\mathrm{AF}}<0 \\
&\varDelta E_{K_{a}}^{\mathrm{AF}}+\varDelta E_{K_{b}}^{\mathrm{AF}}>0
\end{align}
for $0.0 \leq U/t < 3.0$ at $U'/t=0$, indicating that the gain in potential energy leads 
to the staggered CSAF order.  For $3.5 \leq U/t \leq 8.0$, however, the signs are inverted 
and we find that 
\begin{align}
&\varDelta E_{U_{ab}}^{\mathrm{AF}}>0 \\
&\varDelta E_{K_{a}}^{\mathrm{AF}}+\varDelta E_{K_{b}}^{\mathrm{AF}}<0,
\end{align}
which indicates that the gain in kinetic energy leads to the staggered CSAF order.  
Thus, the stabilization mechanism of the CSAF phase also shows a crossover from the Slater 
mechanism to the Mott mechanism.  
With increasing $U'/t$, the $c$-component-related quantities 
$\varDelta E_{U_{bc}}^{\mathrm{AF}}+\varDelta E_{U_{ca}}^{\mathrm{AF}}$ and 
$\varDelta E_{K_{c}}^{\mathrm{AF}}$ vary considerably as shown in Figs.~\ref{fig5}(e) and \ref{fig5}(f).  
We then find that $\varDelta E_{U_{bc}}^{\mathrm{AF}}+\varDelta E_{U_{ca}}^{\mathrm{AF}}>0$ and 
$\varDelta E_{K_{c}}^{\mathrm{AF}}<0$ at $U'\simeq U$, indicating that the $c$-component-related 
quantity loses its potential energy.  However, we find that the $a$- and $b$-component-related 
quantity $\varDelta E_{U_{ab}}^{\mathrm{AF}}<0$ gains considerable potential energy, leading to 
the Slater mechanism of the CSAF ordering.  
At $U \gg U'$, on the other hand, we find that the gain in the kinetic energy of the $a$- 
and $b$-components is dominant, $\varDelta E_{K_{a}}^{\mathrm{AF}}+\varDelta E_{K_{b}}^{\mathrm{AF}}<0$, 
leading to the Mott mechanism of the CSAF ordering.  
Note that the point where the signs of $\varDelta E_{U_{\alpha\beta}}^{\mathrm{AF}}$ and 
$\varDelta E_{K_{\alpha\beta}}^{\mathrm{AF}}$ are inverted approaches $U=U'$ with 
increasing $U'/t$, but the Slater mechanism of the CSAF stabilization even occurs at 
large $U=U'$ values [see Fig.~\ref{fig6}(b)], suggesting that the effects of Fermi surface 
nesting are important around the SU(3) symmetric point, as in the case of the CDW stabilization.  

We thus find that the Slater mechanism (Mott mechanism) of the staggered ordering 
predominantly occurs in the metallic (Mott insulating) region of the paramagnetic phase 
diagram given in Fig.~\ref{fig2}.  
In particular, the Slater mechanism even occurs in the strong coupling region when 
$U'\simeq U$, which is in contrast to the SU(2) symmetric Hubbard model.   
We also find that the fermionic components that show staggered orderings depend on 
the anisotropy of the interaction strengths.  

\begin{figure}[thb]
\begin{center}
\includegraphics[width=\columnwidth]{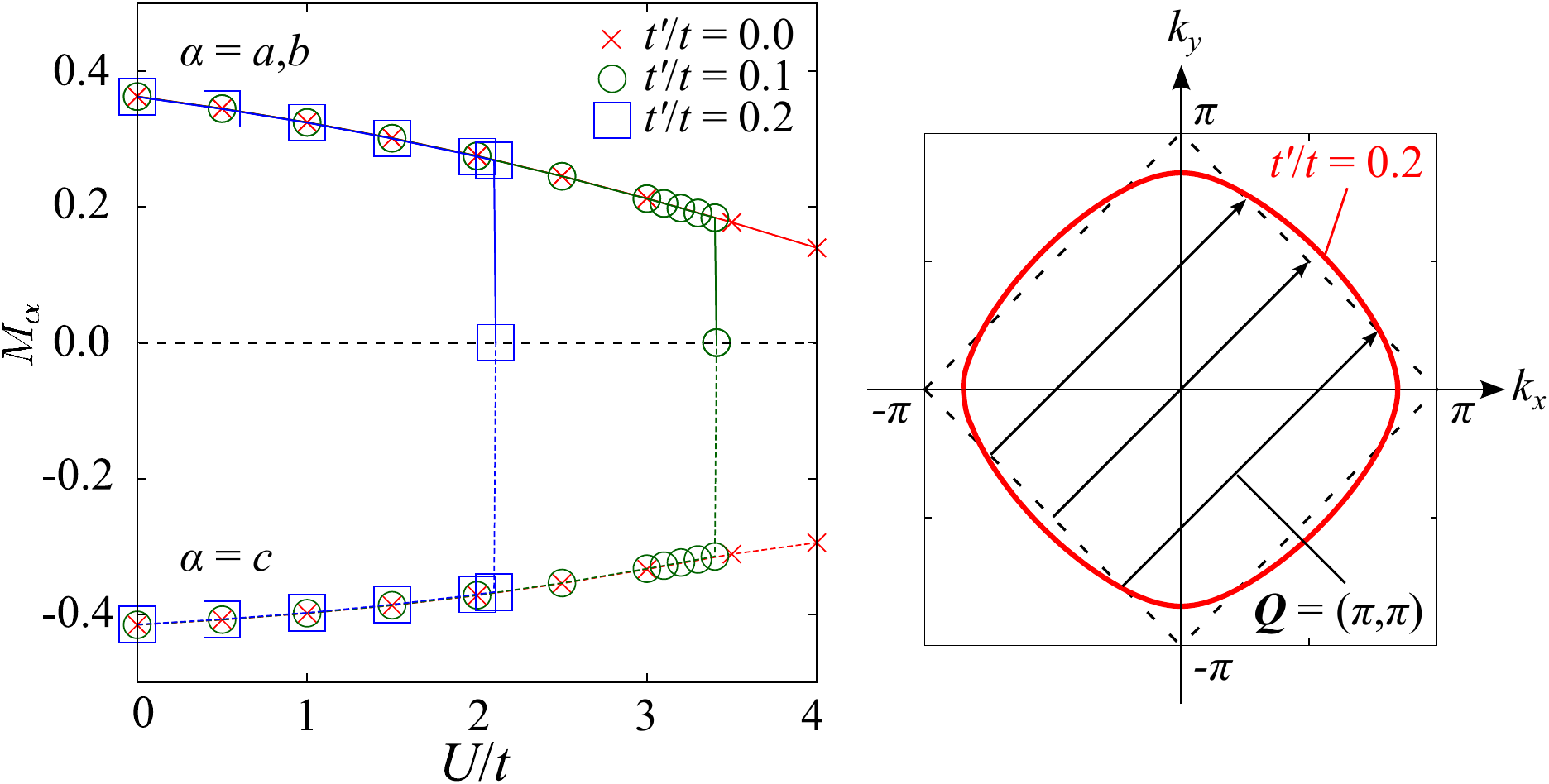}
\caption{(Color online) 
Left panel: Calculated order parameter of the CDW phase $M_\alpha$ in the presence of the 
next-nearest-neighbor hopping $t'$.  The $U/t$ dependence is shown at $U'/t=4$.  
Right panel: Noninteracting Fermi surfaces at $t'/t=0$ and $0.2$.  The nesting vector 
at $t'/t=0$ is indicated by arrows.  
}\label{fig7}
\end{center}
\end{figure}

\subsection{Effects of Fermi surface nesting}

Finally, let us discuss the effects of Fermi surface nesting on the staggered orders.  
In the previous subsection, we showed that the Slater mechanism for the 
stabilization of the staggered orderings occurs in the region of the isotropic interaction 
strengths around the SU(3) symmetric point, where we may expect that Fermi surface 
nesting plays an essential role in the formation of the staggered orders.  
Here, we confirm this expectation by introducing the next-nearest-neighbor hopping term 
to the Hamiltonian and destroying the Fermi surface nesting of $\bm{Q}=(\pi,\pi)$.  
The Hamiltonian then reads 
\begin{align}
{\cal H}=
&- t\sum_{\langle i,j \rangle }\sum_{\alpha} c_{i\alpha}^{\dagger}c_{j\alpha} 
 - t'\sum_{\langle \langle  i,j \rangle \rangle}\sum_{\alpha} c_{i\alpha}^{\dagger}c_{j\alpha} \notag \\
&- \sum_i\sum_{\alpha}\mu_{\alpha}n_{i\alpha}
 + \frac{1}{2}\sum_i\sum_{\alpha\ne\beta}U_{\alpha\beta}n_{i\alpha}n_{i\beta},
\end{align}
where $t'$ is the next-nearest-neighbor hopping parameter and 
$\langle \langle i,j \rangle \rangle$ indicates the summation over the next-nearest-neighbor 
pairs of sites.  The Fermi surface of the noninteracting system at $t'/t=0.2$ is shown 
in Fig.~\ref{fig7}, where we confirm that the nesting feature is completely destroyed.  
We employ the VCA to calculate the CDW order, where the Hamiltonian of the reference system is 
given by $\mathcal{H}'=\mathcal{H}+\mathcal{H}'_{\mathrm{on}}+\mathcal{H}'_{\mathrm{CDW}}$ 
with the on-site potential $\mathcal{H}'_{\mathrm{on}}=\sum_{i,\alpha} \epsilon'_{\alpha}n_{i\alpha}$.  
We optimize the grand potential $\Omega$ with respect to both $M'_{\mathrm{CDW}}$ and $\epsilon'_{\alpha}$; 
the latter is necessary to keep the average particle number at $n_{\alpha}=0.5$ because 
the particle-hole symmetry is broken in this system due to the introduction of the $t'$ term \cite{AAP06}.  

The calculated results for the order parameter $M_\alpha$ are shown in Fig.~\ref{fig7}, 
where we find that the region with a nonvanishing order parameter shrinks with 
increasing $t'/t$ and that the paramagnetic phase without the CDW order actually appears 
around $U=U'$.  We thus demonstrate that, in the region of $U\simeq U'$, the Fermi surface 
nesting of $\bm{Q}=(\pi,\pi)$ plays an essential role in the formation of the staggered 
CDW order, which is not important when $U\ll U'$.  The Slater mechanism thus has a 
contrasting effect to the Mott mechanism for the stabilization of the staggered orders.   

\section{Summary}

We have investigated the ground-state properties and excitation spectra of the 
three-component fermionic Hubbard model defined on the two-dimensional square 
lattice at half filling.  We used the VCA, which enables us to study the effects 
of the lattice geometry and Fermi surface topology in the low-dimensional systems 
in the thermodynamic limit, precisely taking into account spatial fermion correlations.  

First, we presented the ground-state phase diagram of the paramagnetic state 
of the model, whose phases include the paired Mott (PM) phase at $U\ll U'$, the 
color-selective Mott (CSM) phase at $U\gg U'$, and the paramagnetic metallic phase 
between them.  We also showed that the Mott transition does not occur in the SU(3) 
symmetric point $U=U'$ and that a different CSM phase appears between the 
PM and paramagnetic metallic phases, where the color $a$ and $b$ fermions are 
metallic and color $c$ fermions are localized.  

Next, we introduced the Weiss fields to find the spontaneous symmetry-broken 
phases and found that the color-density-wave (CDW) and color-selective 
antiferromagnetic (CSAF) phases appear at $U<U'$ and $U>U'$, respectively, and 
that the energies of the two phases cross at $U=U'$.  
We also examined the kinetic and potential energy gains in the staggered 
orderings and showed that the Slater mechanism with a predominant potential energy 
gain occurs in the region around $U=U'$, where the metallic state is realized in the 
paramagnetic phase, and that the Mott mechanism with a predominant kinetic energy 
gain occurs in the region where the interactions are highly anisotropic and the Mott 
insulating state is realized in the paramagnetic phase.  

By introducing the next-nearest-neighbor hopping parameters, we demonstrated that 
the Fermi surface nesting is essential in the region around $U=U'$, where the Slater 
mechanism occurs for the staggered orderings.  This result indicates that the staggered orders 
near the SU(3) symmetric point are fragile, protected only by the Fermi surface nesting.  
A recent DMFT calculation has suggested that the $s$-wave superfluid state occurs as 
a metastable state at $U<U'$ in the three-component Hubbard model at half filling \cite{OTK14}.  
We may therefore suggest that this superfluid state can be most stable if the Fermi surface nesting 
is destroyed to suppress the CDW order because the pairing of two fermions at $\bm{k}$ 
and $-\bm{k}$ for the superfluidity is not affected strongly by the Fermi surface nesting.  
The exotic pairing mechanism for superfluidity in multicomponent fermionic systems of 
$N>2$ may be an intriguing issue for future studies.  

\section*{Acknowledgments}
We thank A. Koga for enlightening discussions.  
This work was supported in part by KAKENHI Grant No.~26400349 from JSPS of Japan.  
T.~K.~and S.~M.~acknowledge support from the JSPS Research Fellowship for Young Scientists.


\begin{thebibliography}{10}

\bibitem{TN00}
Y.~Tokura and N.~Nagaosa, Science {\bf 288},  462  (2000).

\bibitem{KKRS04} 
A.~Koga, N.~Kawakami, T.~M. Rice, and M.~Sigrist, Phys. Rev. Lett. \textbf{92}, 216402 (2004).  

\bibitem{CLLPM12}
P.~Corboz, M.~Lajk\'o, A.~M. La\"uchli, K.~Penc, and F.~Mila, Phys. Rev. X \textbf{2}, 041013 (2012).  

\bibitem{AM88}
I.~Affleck and J.~B. Marston, Phys. Rev. B {\bf 37},  3774  (1988).

\bibitem{Bl05}
I.~Bloch, Nat. Phys. {\bf 1},  23  (2005).

\bibitem{HGR09}
M.~Hermele, V.~Gurarie, and A.~M. Rey, Phys. Rev. Lett. {\bf 103},  135301
  (2009).

\bibitem{CHU09}
M.~A. Cazalilla, A.~F. Ho, and M.~Ueda, New J. Phys. {\bf 11},  103033  (2009).

\bibitem{GHG10}
A.~Gorshkov, M.~Hermele, V.~Gurarie, C.~Xu, P.~Julienne, J.~Ye, P.~Zoller,
  E.~Demler, M.~Lukin, and A.~Rey, Nat. Phys. {\bf 6},  289  (2010).

\bibitem{NMCLT13} 
H.~Nonne, M.~Moliner, S.~Capponi, P.~Lecheminant, and K.~Totsuka, Europhys. Lett. \textbf{102}, 37008 (2013). 

\bibitem{FTK07}
T.~Fukuhara, Y.~Takasu, M.~Kumakura, and Y.~Takahashi, Phys. Rev. Lett. {\bf
  98},  030401  (2007).

\bibitem{TYS12}
S.~Taie, R.~Yamazaki, S.~Sugawa, and Y.~Takahashi, Nat. Phys. {\bf 8},  825
  (2012).

\bibitem{TTS10}
S.~Taie, Y.~Takasu, S.~Sugawa, R.~Yamazaki, T.~Tsujimoto, R.~Murakami, and
  Y.~Takahashi, Phys. Rev. Lett. {\bf 105},  190401  (2010).

\bibitem{DYM10}
B.~J. DeSalvo, M.~Yan, P.~G. Mickelson, Y.~N. M.~de Escobar, and T.~C. Killian,
  Phys. Rev. Lett. {\bf 105},  030402  (2010).

\bibitem{ZBB14}
X.~Zhang, M.~Bishof, S.~Bromley, C.~Kraus, M.~Safronova, P.~Zoller, A.~Rey, and
  J.~Ye, Science {\bf 345},  1467  (2014).

\bibitem{OLK08}
T.~B. Ottenstein, T.~Lompe, M.~Kohnen, A.~N. Wenz, and S.~Jochim, Phys. Rev.
  Lett. {\bf 101},  203202  (2008).

\bibitem{HWH09}
J.~H. Huckans, J.~R. Williams, E.~L. Hazlett, R.~W. Stites, and K.~M. O'Hara,
  Phys. Rev. Lett. {\bf 102},  165302  (2009).

\bibitem{GB09}
E.~V. Gorelik and N.~Bl\"umer, Phys. Rev. A {\bf 80},  051602  (2009).

\bibitem{IMS10}
K.~Inaba, S.~Miyatake, and S.~Suga, Phys. Rev. A {\bf 82},  051602(R)  (2010).

\bibitem{IS13}
K.~Inaba and S.~Suga, Mod. Phys. Lett. B {\bf 27},  1330008  (2013).

\bibitem{MIS10}
S.~Miyatake, K.~Inaba, and S.~Suga, Phys. Rev. A {\bf 81},  021603(R)  (2010).

\bibitem{YK16}
H.~Yanatori and A.~Koga, J. Phys. Soc. Jpn. {\bf 85},  014002  (2016).

\bibitem{YK161}
H.~Yanatori and A.~Koga, eprint arXiv:1603.02647.

\bibitem{IS12}
K.~Inaba and S.~Suga, Phys. Rev. Lett. {\bf 108},  255301  (2012).

\bibitem{OTK14}
Y.~Okanami, N.~Takemori, and A.~Koga, Phys. Rev. A {\bf 89},  053622  (2014).

\bibitem{Su15}
S.~Suga, Phys. Rev. A {\bf 92},  023617  (2015).

\bibitem{HH04}
C.~Honerkamp and W.~Hofstetter, Phys. Rev. Lett. {\bf 92},  170403  (2004).

\bibitem{HH04-2}
C.~Honerkamp and W.~Hofstetter, Phys. Rev. B {\bf 70},  094521  (2004).

\bibitem{RZH07}
A.~Rapp, G.~Zar\'and, C.~Honerkamp, and W.~Hofstetter, Phys. Rev. Lett. {\bf
  98},  160405  (2007).

\bibitem{RHZ08}
A.~Rapp, W.~Hofstetter, and G.~Zar\'and, Phys. Rev. B {\bf 77},  144520
  (2008).

\bibitem{IS09}
K.~Inaba and S.~Suga, Phys. Rev. A {\bf 80},  041602(R)  (2009).

\bibitem{SK14}
M.~Sakaida and N.~Kawakami, Phys. Rev. A {\bf 90},  013632  (2014).

\bibitem{PAD03}
M.~Potthoff, M.~Aichhorn, and C.~Dahnken, Phys. Rev. Lett. {\bf 91},  206402
  (2003).

\bibitem{Po12}
M.~Potthoff,  in {\em Strongly Correlated Systems -- Theoretical Methods},
  Vol.~171 of {\em Springer Series in Solid-State Sciences} (Springer-Verlag,
  Berlin Heidelberg, 2012), Chap.~10, p.~303.

\bibitem{Po03-1}
M.~Potthoff, Eur. Phys. J. B {\bf 32},  429  (2003).

\bibitem{Po03-2}
M.~Potthoff, Eur. Phys. J. B {\bf 36},  335  (2003).

\bibitem{SLM05} 
D.~S\'en\'echal, P.-L.~Lavertu, M.-A.~Marois, and A.-M.~S.~Tremblay, 
Phys. Rev. Lett. \textbf{94}, 156404 (2005).  

\bibitem{ST04} 
D.~S\'en\'echal and A.-M.~S.~Tremblay, Phys. Rev. Lett. \textbf{92}, 126401 (2004).  

\bibitem{DAH04}
C.~Dahnken, M.~Aichhorn, W.~Hanke, E.~Arrigoni, and M.~Potthoff, Phys. Rev. B
  {\bf 70},  245110  (2004).

\bibitem{SPP00}
D.~S\'en\'echal, D.~Perez, and M.~Pioro-Ladri\`ere, Phys. Rev. Lett. {\bf 84},
  522  (2000).

\bibitem{SPP02} 
D.~S\'en\'echal, D.~Perez, and D.~Plouffe, Phys. Rev. B {\bf 66}, 075129  (2002).

\bibitem{Se12} 
D.~S\'en\'echal, in {\em Strongly Correlated Systems --Theoretical Methods}, 
Vol.~171 of {\em Springer Series in Solid-State Sciences} (Springer-Verlag, 
Berlin Heidelberg, 2012), Chap.~8, p.~237.  

\bibitem{YOT13}
H.~Yokoyama, M.~Ogata, Y.~Tanaka, K.~Kobayashi, and H.~Tsuchiura, J. Phys. Soc.
  Jpn. {\bf 82},  014707  (2013).

\bibitem{KO14}
T.~Kaneko and Y.~Ohta, J. Phys. Soc. Jpn. {\bf 83},  024711  (2014).

\bibitem{HSY14}
H.~Watanabe, T.~Shirakawa, and S.~Yunoki, Phys. Rev. B {\bf 89},  165115
  (2014).

\bibitem{AAP06}
M.~Aichhorn, E.~Arrigoni, M.~Potthoff, and W.~Hanke, Phys. Rev. B {\bf 74},
  024508  (2006).

\end{thebibliography}
\end{document}